\documentstyle[12pt]{article}
\begin{document}

\centerline{\bf Three-Species Diffusion-Limited Reaction with}
\centerline{\bf Continuous Density-Decay Exponents}
\centerline{$\;$}
\centerline{Jae Woo Lee$^{1,2}$ and Vladimir Privman$^{2}$}
\centerline{$\;$}
\noindent{$^{1}$Department of Physics, Inha University,
Inchon 402-751, Korea}
\centerline{$\;$}
\noindent{$^{2}$Department of Physics, Clarkson University, 
Potsdam, NY 13699-5820, USA}

\begin{abstract}
We introduce a model of three-species
two-particle diffusion-limited reactions $A+B
\rightarrow A$ or $B$, $B+C \rightarrow B$ or $C$, and $C+A \rightarrow
C$ or $A$, with three persistence parameters (survival probabilities
in reaction) of the hopping particle. We consider isotropic and 
anisotropic diffusion (hopping with a drift) in $1d$.
We find that the particle density decays as a power-law for certain
choices of the
persistence parameter values. In the anisotropic case, on
one symmetric line in the parameter space, the decay 
exponent is {\it monotonically
varying\/} between the values close to 1/3 and 1/2.
On another, less symmetric line, the exponent is constant. For
most parameter values, the density does not follow a
power-law. We also calculated various characteristic exponents for
the distance of nearest particles and domain
structure. Our results support
the recently proposed possibility that $1d$
diffusion-limited reactions with a drift do not fall
within a limited number of distinct universality classes.\\
\end{abstract}

\ 

\ 

\noindent {\large J. Phys. A {\bf 30}, L317-L324 (1997)}

\ 

\ 


\newpage

The kinetics of diffusion-limited reactions has been extensively
studied, with recent emphasis on fluctuations in low-dimensional
systems [1-7].
In reactions with symmetric annihilation or coagulation of species,
including symmetric initial conditions, the density follows a
power-law $C(t) \sim t^{-\alpha}$, with a nontrivial critical exponent
$\alpha$ below the upper critical
dimension $d_c$; see [3,5-9].
For instance, for single-species annihilation $A+A \rightarrow 0$
and coagulation
$A+A \rightarrow A$, the density decays asymptotically according to the
power-law $C(t) \sim t^{-1/2}$ for $d =1 <
d_c=2$, and according to the mean-field power-law
$C(t) \sim t^{-1}$ for $ d> d_c$, etc.; see [10-19].

For two-species annihilation $A+B \rightarrow 0$ the density follows
the power-law decay $C(t) \sim t^{-d/4} $ for $d < d_c=4$
and $C(t) \sim t^{-1}$ (mean-field) for $d>4$
[8,20-24]. Recently, it was found
that in the two-species annihilation model with hard-core particle
interactions (same-species exclusion) 
in $d=1$ the drift in
particle hopping changes the critical exponent $\alpha$
from $1/4$ to $1/3$ [25-27].

There are also several studies of multiparticle reactions such as $kA
\rightarrow 0$ or $A+B+C \rightarrow 0$ [28-35]. For instance, 
for $kA \rightarrow 0$ the upper critical dimension is
$d_c=2/(k-1)$ [28,30-32].
For $d>d_c$, the system follows the mean-field rate equation $dC/dt
\sim - C^k$ and the density decays as $C(t) \sim t^{-1/(k-1)}$. For $d<
d_c$, the fluctuations are important, while at $d_c$, logarithmic
corrections are generally expected in the mean-field power-laws.
The general reaction $A_1 +A_2 \cdots +A_k \rightarrow 0$ has
also been studied by scaling arguments [9,22,23].

Recently, a model was introduced of diffusion-limited
reactions of two
species of particles, $A+B \rightarrow A$ or $B$, with a drift
in diffusion and with hard-core interactions between same-species
particles. The decay exponent of
the density was found to vary {\it continuously\/} as a function
of the probability of
which particle, the hopping one or the target, survives in the
reaction [36]. This study has suggested that diffusion-limited
reactions with drift (anisotropy) in
the diffusion of particles do not fall
within few distinct universality classes in $d=1$.

In the present work, we extend this observation to a
three-species hard-core two-particle reaction model in $1d$.
Our model has three adjustable parameters, the
survival probabilities, in reaction, of the hopping particle. All
our results
were obtained by extensive numerical Monte Carlo simulations
utilizing concurrently a cluster of over 50 IBM RISC-6000
workstations at Clarkson University.
In the rest of this work, we first define the model generally, and
then report numerical results for various parameter values.

Our model is an extension of the two-species model on the $1d$
lattice. Each lattice site can be occupied by a single
particle ($A$ or $B$ or $C$), or empty. Monte Carlo simulations
were performed
for the cases of isotropic and maximally anisotropic hopping.
In the isotropic case, a randomly selected particle
attempts to hop to the left or  right nearest-neighbor site
with equal probabilities $1/2$. In the general 
anisotropic case, the particle attempts to hop
to the right with probability
$(1+a)/2$ or to the left with probability $(1-a)/2$. In this work 
we took the maximal bias $a=1$, i.e., the chosen particle only
attempts to hop to the right. 

If the target site is empty then the hopping attempt succeeds and
the chosen particle is moved one lattice spacing. If the target
site is occupied by a particle of the same species as the chosen
particle then the hopping attempt fails; this rule models hard-core
interaction between same-species particles. If the target site is
occupied by a
particle of a different species then the hopping is accompanied by
reaction defined by the following probabilistic rules (shown here for
hopping to the right):
\begin{center}
\begin{equation}
AB \rightarrow \left\{ \begin{array}{ll}
0A & {\rm Prob.}\; p \\
0B & {\rm Prob.}\; 1-p \\
\end{array}
\right.
{}~{\rm and}~~~
BA \rightarrow \left\{ \begin{array}{ll}
0B & {\rm Prob.}\; p \\
0A & {\rm Prob.}\; 1-p \\
\end{array}
\right.
\end{equation}
\end{center}
\begin{center}
\begin{equation}
BC \rightarrow \left\{ \begin{array}{ll}
0B & {\rm Prob.}\; q \\
0C & {\rm Prob.}\; 1-q \\
\end{array}
\right.
{}~{\rm and}~~~
CB \rightarrow \left\{ \begin{array}{ll}
0C & {\rm Prob.}\; q \\
0B & {\rm Prob.}\; 1-q \\
\end{array}
\right.
\end{equation}
\end{center}
\begin{center}
\begin{equation}
CA \rightarrow \left\{ \begin{array}{ll}
0C & {\rm Prob.}\; r \\
0A & {\rm Prob.}\; 1-r \\
\end{array}
\right.
{}~{\rm and}~~~
AC \rightarrow \left\{ \begin{array}{ll}
0A & {\rm Prob.}\; r \\
0C & {\rm Prob.}\; 1-r \\
\end{array}
\right.
\end{equation}
\end{center}
where $0 \leq p, q, r \leq 1$. The probabilities $p$, $q$, $r$ represent
the persistence (probability of survival) of the hopping particle.
Thus the reactions involved are $A+B \rightarrow A$ or $B$,
$B+C \rightarrow B $ or $C$,
and $C+A \rightarrow C$ or $A$. 

In the Monte Carlo simulations, we used lattices of $10^5$
sites with periodic boundary
conditions. One Monte Carlo time step corresponded, statistically, to
the number of hopping attempts equal the number of remaining 
particles, so that, on average, each particle's hopping attempt rate
(per unit time) was 1. Initial
densities were 90\% of the full occupancy, with randomly distributed
equal
densities of the three species. Data were collected for up to $10^5$
Monte Carlo time steps and averaged over at least 100 runs
for each choice of the persistence parameter values. 

Let us point out that if the initial density of the $C$ particles, for
instance, is zero, then our model becomes two-species, identical to
that studied in [36]. In this two-species ($A$ and $B$) case,
the initial symmetry $A \leftrightarrow B$, assuming equal densities,
is maintained dynamically for all values of $p$. Indeed, the number of
$A\cdots B$ configurations 
(here $\cdots$ represent empty or no sites)
which lead to a reaction when $A$ ``catches
up'' with $B$ (we consider the fully anisotropic hopping case
here [36]) is equal, on the $1d$ lattice, to the number of
$B \cdots A$ configurations: they simply alternate.

In the new, three-species model, the symmetries are less robust.
Indeed,
starting from a symmetric initial distribution, the system can evolve
dynamically into a state which is not symmetric with respect to the
three species involved. In this regard, our results shed an
interesting light on the nature of the nonuniversal-exponent
behavior. Similar to the two-species case, we find nonuniversal
exponents only when all the following conditions are satisfied, and
presumably it is the interplay of all three of them that leads to
nonuniversality: the hopping must be anisotropic, the same-species
interaction must be hard-core, and the {\it full symmetry\/} must be
maintained. Thus, we find nonuniversal exponents only on the
symmetric line $p=q=r$ in the parameter space. On some other lines, we
find constant-exponent (universal) behavior, while in most of the
parameter space, lack of symmetry results in a non-power-law density
variation (so that critical exponents are not defined).
We note, however, that for a certain three-species system in $2d$,
with {\it three-particle reactions}, continuous exponents were found
[37] for a specific line in the parameter space, without introduction
of spatial anisotropy in the dynamics.

In Fig.~1, we plot the density as a function of time
for varying persistence parameter values, for the fully
symmetric case $p=q=r$ and anisotropic hopping.
The log-log plot clearly shows the asymptotic power-law
(straight-line) behavior. However, the slope depends on the
persistence parameter. The decay exponents were estimated by
extrapolation of the local slopes, similar to [36].
In Table~1, we list the exponents for various values of the persistence
parameter. We actually calculated several physical quantities which
characterize fluctuations in the system, similar to [36]. These
include the density 
$C(t) \sim t^{-\alpha}$, the average distance between nearest particles
of the same species $\langle l_{AA} (t) \rangle \sim t^{\beta}$, the
average
distance between nearest particles of different species
$\langle l_{AB} (t) \rangle
\sim t^{\gamma}$, the average domain size of same-species
particles $ \langle L_A(t) \rangle
\sim t^{\delta}$, the average number of particles per such domain 
$ \langle N_A(t) \rangle
\sim t^{\eta}$, the average number of pairs of same-species
particles $\langle N_{AA}
(t) \rangle \sim t^{-\mu}$, and the average number of  
pairs of different-species
particles $\langle N_{AB} (t) \rangle \sim t^{-\nu}$. In the latter two
quantities the pairs need not be nearest-neighbor, they can be separated
by empty lattice sites. All these
exponents vary nonuniversally along the line $p=q=r$; their values
will be further discussed in the following paragraphs.

The isotropic-hopping case was studied numerically only for $p=q=r$
in this work. We found that the exponents do not depend on
the persistence parameter. The characteristic exponents for the
isotropic
hopping were estimated as $\alpha=0.350(5)$, $\beta=0.366(5)$, $\gamma
=0.438(8) $, $\delta=0.496(4) $, $\eta=0.164(8)$, $\mu=0.328(3)$, and
$\nu = 0.512(6)$, where the
uncertainties always refer to the last digit.
The density-decay exponent, $\alpha$, value is somewhat
larger than $1/3$ but too small for consistency with
predictions for
multiparticle annihilation, e.g., $A_1 + A_2 +
A_3 \rightarrow 0$; see [28]. 
Furthermore, it is somewhat smaller than the 
prediction $\alpha =3/8=0.375$ for
three-species two-particle annihilation $A_i + A_j \rightarrow 0 ~~(i
\neq j )$ [32]. As already mentioned, exponent values exactly
(or very close to) $1/3$ appear in the fully anisotropic $A+B\to 0$
reaction. The value $\alpha$ near $1/3$ was also found for the
two-species variant of our model with anisotropy and with $p=1/2$, and
in the anisotropic three-species case with $p=q=r=0$ (see Table~1).

Thus, if seems likely that both the isotropic-hopping results and the
anisotropic results, with the latter limited to certain special
points in the parameter space,
will be eventually identified within some established
universality classes of various diffusion-limited reactions. 

However, for anisotropic hopping, the exponents are nonuniversal when
the persistence parameter values are varied in the full
range from 0 to 1,
both in the symmetric three-species case and in the
two-species case (where the symmetry is built-in).  Of course,
there is always the danger that the observed behavior, interpreted as
nonuniversality, as actually a slow crossover phenomenon. However, we
note that our simulation is sufficiently ``large-scale'' as compared
to other simulations (including our own in this work) which have found
both universal and nonuniversal behavior. So, we feel confident that
the observed nonuniversality is well-established within the limits of
modern computational capabilities.

Some general
exponent properties can still be discussed even if the universality
class association is ambiguous. Owing to the
effective repulsion, one expects that $\langle l_{AA} (t) \rangle
\leq \langle l_{AB} (t) \rangle $,
i.e., $\beta \leq \gamma$. This inequality is always satisfied by our
results (including the case of less symmetry discussed later). 
There is another inequality, $\beta \geq \alpha$, which holds
because the average
interparticle  distance is related to the (fluctuating)
particle density $c_A (t)$ according to
$\langle l_{AA} \rangle = \langle 1/c_A (t) \rangle \geq 1/
\langle c_A(t) \rangle$ [38,39]. Our
simulation results also satisfy this relation within error bars.

Since the density is approximately equal to the number of particles per
domain divided by the average domain size, i.e., $C_A
\sim \langle N_A \rangle / \langle L_A \rangle
$ [36,39], the exponents $\delta $ and $\eta$ should
satisfy the relation $\alpha=
\delta - \eta$. Our data are consistent with this relation.
The rate of change of the $A$-particle density is
proportional to the number of the pairs $A\cdots B$,
$\langle N_{AB} \rangle$,
divided by the diffusion time which is of
order $\langle\ell_{AB}\rangle^2/
\cal{D}$, where $\cal{D}$ is the diffusion constant. This yields
the exponent relation $\alpha=2\gamma+\nu-1$ [39]. Another
exponent relation
follows by observing that the same rate can be estimated as the inverse
of $\langle L_A \rangle \langle\ell_{AB}\rangle^2/
\cal{D}$, which yields $\alpha=2\gamma+\delta-1$ [26]. Combining
the above relations, we get $\nu=\delta$ and $2\gamma+\eta=1$. The
latter equalities are satisfied by our results to within 10\%. Note
that these exponent relations are based on a combination of mean-field
and diffusive arguments and they are therefore phenomenological.

For the symmetric dynamics with $p=q=r=0$ and anisotropic
hopping, the exponents are equal to those for
the isotropic hopping case within error bars. The estimated exponent of
the density is close to $\alpha =1/3$. When values of the persistence
parameters increase along the diagonal (symmetric)
line in the parameter space the
exponent of the density $\alpha$ increases continuously in the
anisotropic case. At $p=q=r=1$ the
density exponent estimate is close to $\alpha=1/2$ (the latter value is
likely exact). The decay of the
density $\sim t^{-1/2}$ is then similar to that of the single-species
coalescence or annihilation,
$A+A \rightarrow A$ or $0$, and the two-species annihilation
version of our model [36], discussed earlier, at $p=1$. 
In this case no large domains
of the same species of particle are formed. Indeed,
the exponent $\eta$ estimates are close to zero. Even in this well-mixed
situation, non-mean-field fluctuations can arise in the form of
non-mean-field interparticle distribution [12,34].
For $p=q=r=1$, the ``catching up'' argument for the
impossibility of large same-species domains applies, similarly to the
two-species case. This argument is not reviewed here; see [36].

For anisotropic hopping, we explored points in the parameter
space of $p,q,r$ outside the symmetric line $p=q=r$.
The only other regions in the parameter space where power-law behavior
is found are the line $p=q=1$, $r$-varying, and two lines
obtained by relabeling.
In Fig.~2, we plot the density versus time 
for $p=q=1$ and several $r$ values. For large time, power-law behavior
is obtained
with the same exponents for $r<1$ as in the case $p=q=r=1$, within
error limits. It is important to recall that for $p=q=r=1$ the system
is fully mixed: there are no large same-species domains formed. For
$r<1$, the $B$ species is no longer symmetric (while the $A
\leftrightarrow C$ symmetry is 
still preserved). Numerical indications are that the density of $B$
still follows approximately the same power law as $A$ and $C$,
see Fig.~3 for the case $r=0$, but with a smaller
amplitude. The well-mixed state seems to persist for $r<1$. 

As already mentioned, probes at several other $p,q,r$ values with
varying degree of symmetry (though we did not do a ``dense'' scan of the
full cube $0 \leq p,q,r \leq 1$) 
seem to suggest that outside the lines $p=q=r$, $p=q=1$
(and also $p=r=1$
and $q=r=1$ by symmetric relabeling), the behavior of the density is
no longer power-law.
Let us consider, for illustration, the point $p=1$, $q=r=0$. The
symmetry here 
seems not lower than, for instance, the line $q=r=1$. In both cases,
$C$ is special while $A$ and $B$ remain symmetric. However, for
$p=1$, $q=r=0$, our numerical data suggest that $C$ particles survive
with nonzero final density, see Fig.~4, while $A$ and $B$ are
eliminated faster than power-law for large times, as shown in Fig.~4.
Attempts to fit the $A$
density to a stretched exponential were inconclusive (the fitted
exponent of the stretched-exponential power was
very small). Generally, for 
$p,q,r$ values which are not symmetrically positioned
in the parameter space 
there is no reason to expect equal large-time densities of
the species even 
for equal-density initial conditions.

Finally, let us list some preliminary findings [40] which
hopefully illuminate the robustness of the results of this work to 
changes in the reaction rules. Numerical Monte Carlo simulations
[40] suggest that the following symmetric three-species reaction,
$A+B \to C$, $B+C \to A$,
$C+A \to B$, has the critical exponent $\alpha={1 \over 2}$ regardless
of the drift. However, for the three-species two-particle annihilation
reaction, $A+B \to 0$, $B+C \to 0$, $C+A \to 0$, the exponents seems 
to depend on the drift [40].

In summary, we studied three-species diffusion-limited reactions with
emphasis on the effects of hopping anisotropy and variation of the
survival probabilities of the hopping particles. The ``critical''
power-law behavior was observed only along special, symmetric lines in
the parameter space. In the full three-species symmetry case the
critical exponents vary continuously, with that for the particle
density increasing from about $1/3$ to $1/2$ when $p=q=r$ increase
from zero to one. On the less symmetric line $p=q=1$, the exponents
for varying $r<1$ are the same as for $p=q=r=1$. 

{\sl This work was supported in part by Inha University. This financial
assistance is gratefully acknowledged.}

\newpage
\centerline{\bf References}

\ 

\frenchspacing{

\noindent\hang [1] T. Liggett, ``Interacting Particle Systems''
(Springer-Verlag, NY, 1985).

\noindent\hang [2] V. Privman, Trends Stat. Phys. {\bf 1}, 89 (1994).

\noindent\hang [3] ``Nonequilibrium Statistical Mechanics in One
Dimension,'' ed. V. Privman (Cambridge University Press,
Cambridge, 1997).

\noindent\hang [4] V.
Kuzovkov and E. Kotomin, Rep. Prog. Phys. {\bf 51},
1479 (1988).

\noindent\hang [5] A. A. Ovchinikov and Ya. B. Zeldovich, Chem. Phys.
{\bf 28}, 215 (1978).

\noindent\hang [6] D. Toussaint and F. Wilczek, J. Chem. Phys. {\bf 78},
2642 (1983).

\noindent\hang [7] D. C. Torney and H. M. McConnell, J. Phys. Chem.
{\bf 87}, 1941 (1983). 

\noindent\hang [8] K. Kang and S. Redner, Phys. Rev. Lett. {\bf 52},
955 (1984).

\noindent\hang [9] K. Kang and S. Redner, Phys.
Rev. A {\bf 30}, 2833 (1984).

\noindent\hang [10] V. Privman, A. M. R. Cadilhe and M.L. Glasser,
J. Stat. Phys. {\bf 81}, 881 (1995).

\noindent\hang [11] D. J. Balding and N. J. B.
Green, Phys. Rev. A {\bf 40}, 4585 (1989).

\noindent\hang [12] D. ben-Avraham, M. A. Burschka and C. R. Doering, J.
Stat. Phys. {\bf 60}, 695 (1990).

\noindent\hang [13] M. Bramson and D. Griffeath, Ann. Prob. {\bf 8},
183 (1980).

\noindent\hang [14] A. A. Lushnikov, Phys. Lett. A
{\bf 120}, 135 (1987).

\noindent\hang [15] V. Privman, J. Stat. Phys. {\bf 69}, 629 (1992).

\noindent\hang [16] V. Privman, J. Stat. Phys. {\bf 72}, 845 (1993).

\noindent\hang [17] V. Privman, Phys. Rev. E {\bf 50}, 50 (1994).

\noindent\hang [18] J. C. Lin, Phys. Rev. A {\bf 44}, 6706 (1991).

\noindent\hang [19] J. C. Lin, C. R. Doering and D. ben-Avraham,
Chem. Phys. {\bf 146}, 355 (1990). 

\noindent\hang [20] M. Bramson and J. L. Lebowitz,
Phys. Rev. Lett. {\bf 61}, 2397 (1988).

\noindent\hang [21] M. Bramson and J. L. Lebowitz, J.
Stat. Phys. {\bf 62}, 297 (1991).

\noindent\hang [22] K. Kang and S.
Redner, Phys. Rev. A {\bf 32}, 435 (1985).

\noindent\hang [23] B. P. Lee and J. Cardy, J. Phys. A
{\bf 80}, 971 (1995).

\noindent\hang [24] H. Simon, J. Phys. A {\bf 28}, 6585 (1995).

\noindent\hang [25] S. A. Janowsky, Phys. Rev. E {\bf 51}, 1858 (1995).

\noindent\hang [26] S. A. Janowsky, Phys. Rev. E {\bf 52}, 2535 (1995).

\noindent\hang [27] I. Ispolatov, P. L. Krapivsky and
S. Redner, Phys. Rev. E {\bf 52}, 2540 (1995). 

\noindent\hang [28] K. Kang, P. Meakin, J. H. Oh and S. Redner, J.
Phys. A {\bf 17}, L665 (1994).

\noindent\hang [29] V. Privman and M. D. Grynberg, J. Phys. A
{\bf 25}, 6567 (1992).

\noindent\hang [30] V. Privman, E. Burgos and M. D. Grynberg,
Phys. Rev. E {\bf 52}, 1866 (1995).

\noindent\hang [31] B. P. Lee, J. Phys. A {\bf 27}, 2633(1994).

\noindent\hang [32] D. ben-Avraham and S. Redner, Phys. Rev. A {\bf 34},
501 (1986).

\noindent\hang [33] D. ben-Avraham, Phil. Mag. {\bf 56}, 1015 (1987).

\noindent\hang [34] D. ben-Avraham, Phys. Rev. Lett.
{\bf 71}, 3733 (1993).

\noindent\hang [35] G. Oshanin, A. Stemmer, S.
Luding and A. Blumen, Phys. Rev. E {\bf 52}, 5800 (1995).

\noindent\hang [36] D. ben-Avraham, V. Privman and D.
Zhong, Phys. Rev. E {\bf 52}, 6889 (1995).

\noindent\hang [37] T. J. Newman, J. Phys. A {\bf 28}, L183 (1995).

\noindent\hang [38] F. Leyvraz and S. Redner, Phys. Rev.
Lett. {\bf 66}, 2168 (1991).

\noindent\hang [39] F. Leyvraz and S. Redner, Phys. Rev.
A {\bf 46}, 3132 (1992).

\noindent\hang [40] J. W. Lee, to be published.

}
\newpage
\noindent\hang {\bf Table 1:\ }\ Exponent estimates for
several values of the
persistence parameters, with $p=q=r$. The
exponents are defined according to $C \sim t^{-\alpha}$,
$ \langle l_{AA}
\rangle  \sim t^{\beta}$,
$\langle l_{AB}\rangle \sim t^{\gamma}$,
$\langle L_A\rangle \sim t^{\delta}$, $\langle N_A\rangle \sim
t^{\eta}$, $\langle N_{AA}\rangle \sim t^{-\mu}$,
$\langle N_{AB}\rangle \sim t^{-\nu}$; see text for further details.

\vskip 0.4 true in

\begin{center}
\begin{tabular}{llllllll}
$p$& $\alpha $ & $\beta$ & $\gamma$
& $\delta$ & $\eta$ & $\mu$ & $\nu $
\\
1  & 0.499(2)$\;\;\;$ & 0.49(1) & 0.50(1)
& 0.503(6)$\;\;\;$ & 0.005(4)$\;\;\;$ & 0.497(4)$\;\;\;$
& 0.499(5)$\;\;\;$ \\
0.75$\;\;\;$& 0.456(3) & 0.46(1) &
0.47(1) & 0.51(1) & 0.064(3) & 0.440(8) & 0.525(5) \\
0.5 & 0.402(3) & 0.42(1) & 0.44(1) & 0.54(1) & 0.16(1) & 0.381(7) &
0.558(5) \\
0.25 & 0.360(4) & 0.387(3) & 0.42(1) & 0.56(1) & 0.23(1) & 0.341(2)
&
0.578(4) \\
0 & 0.340(4) & 0.360(2)$\;\;\;$ & 0.405(6)$\;\;\;$ &
0.57(1) & 0.26(1) & 0.316(4) &
0.60(3) 
\end{tabular}
\end{center}
\newpage
\centerline{\bf Figure Captions}

\ 

\noindent\hang Figure 1:
Log-log plot of the decay of the $A$-particle density
versus time, for $p=q=r=1$ (bottom curve), 0.75,
0.5, 0.25, 0 (topmost curve).

\ 

\noindent\hang Figure 2:
Log-log plot of the decay of the $A$-particle density
versus time, for $p=q=1$ and $r=1$ (bottom curve), 0.75, 0.5, 0.25,
0 (topmost curve).

\ 

\noindent\hang Figure 3:
Log-log plot of the decay of
the $A$ (upper curve) and
$B$ (lower curve) particle densities for $p=q=1$,
$r=0$.

\ 

\noindent\hang Figure 4:
Log-log plot of the
variation of the $C$-particle density (upper curve)
and $A$ (and $B$) density (lower curve) versus time, for $p=1$,
$q=r=0$.

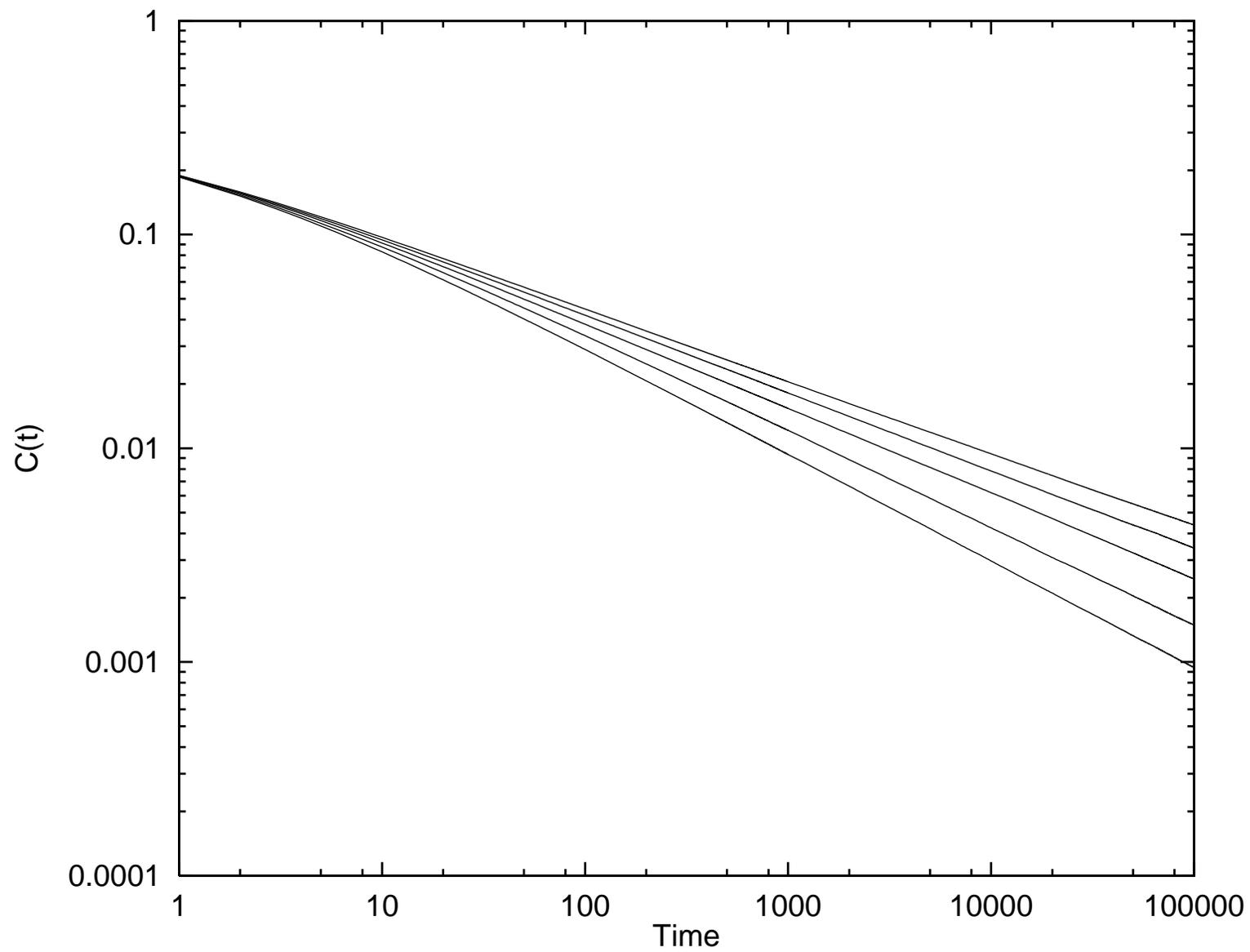
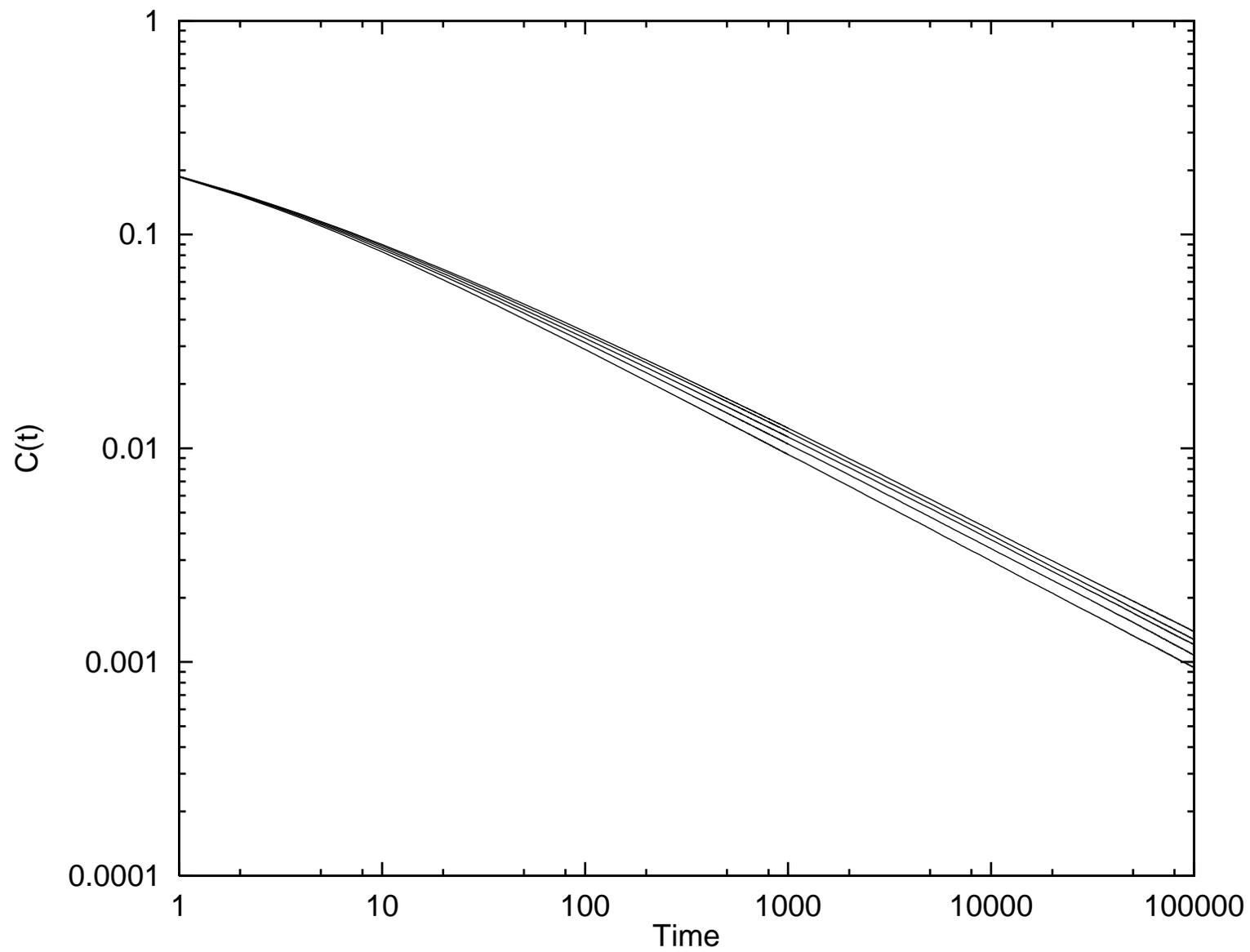
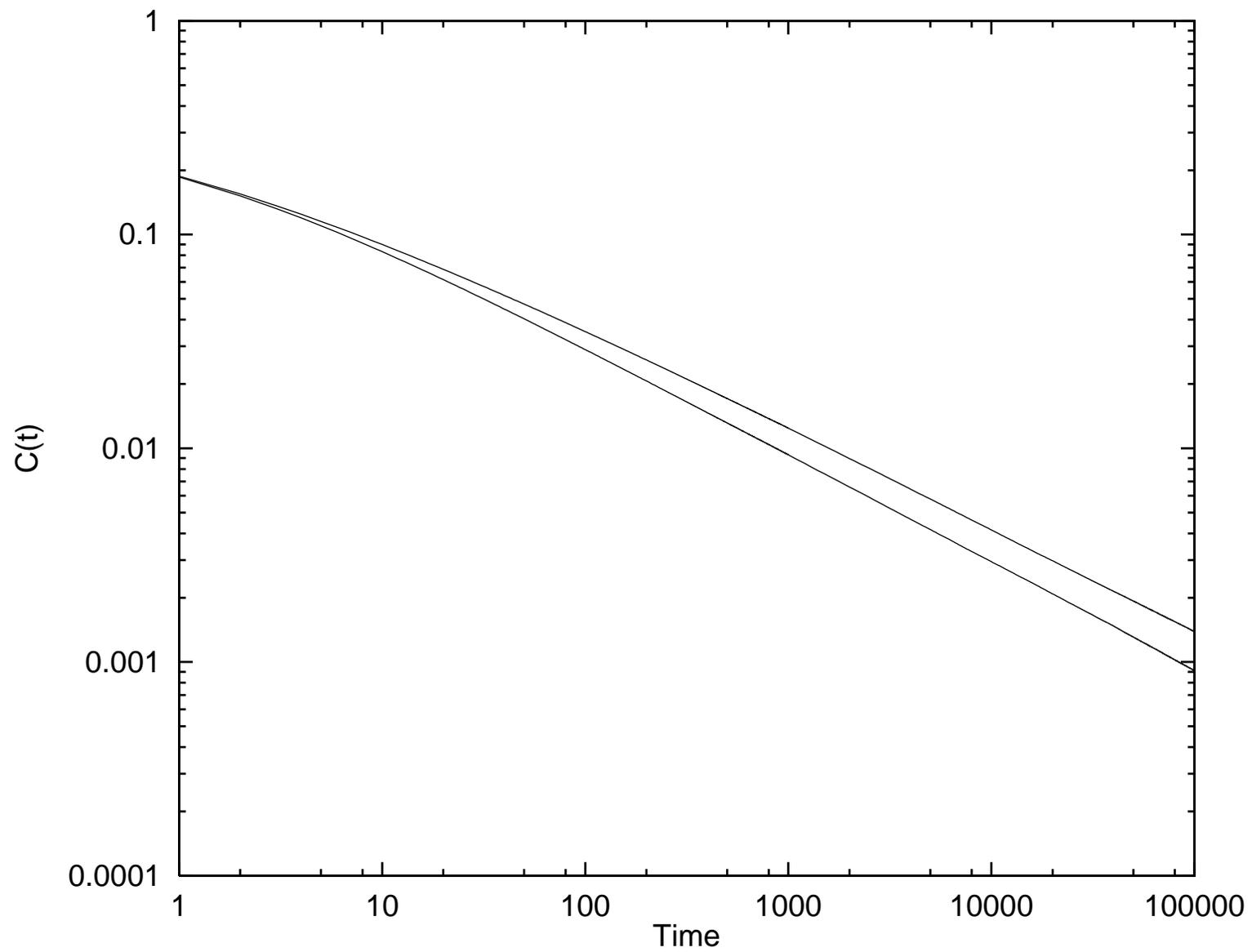
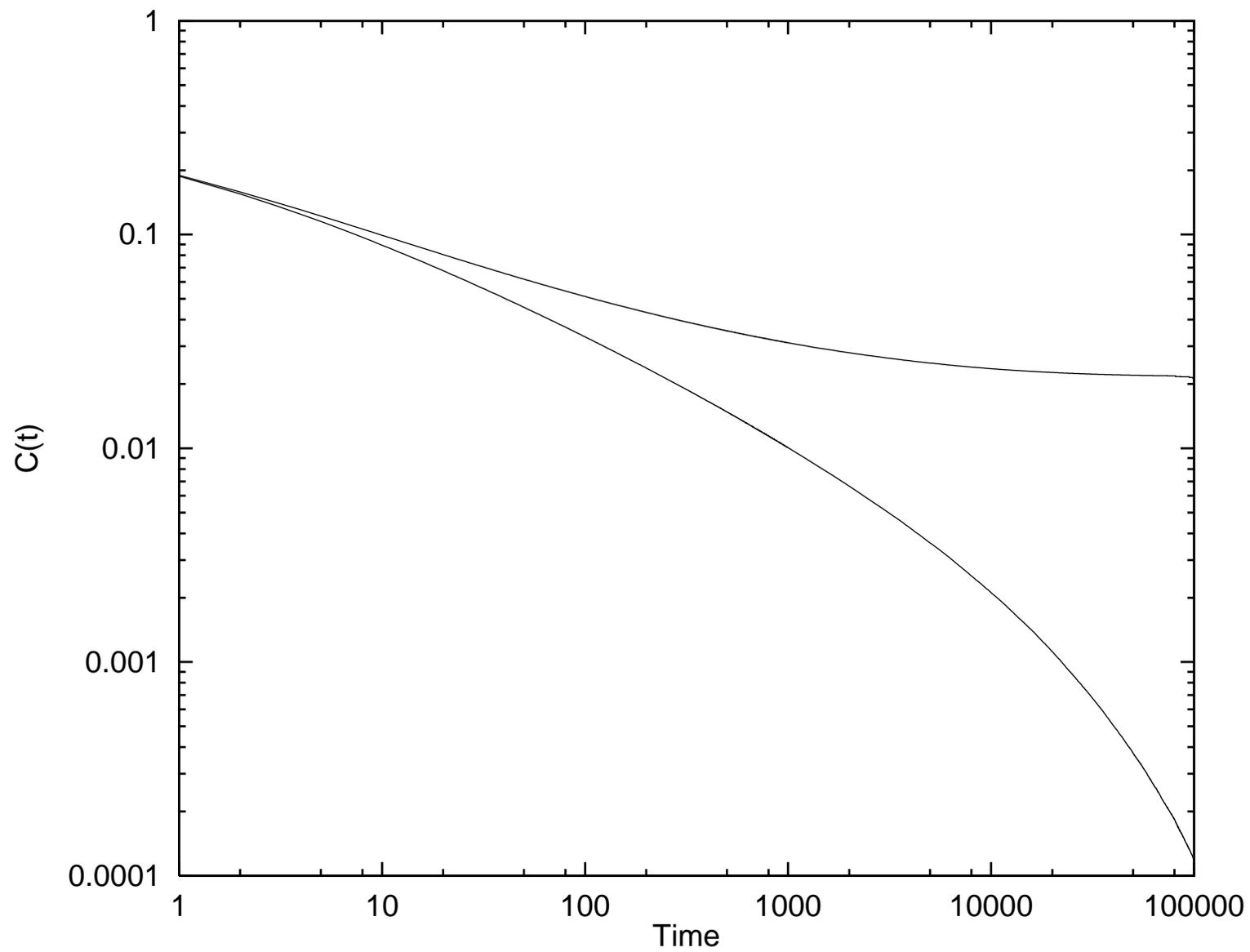
\end{document}